\documentstyle[fleqn]{article}
\def\beq{\begin{equation}}
\def\eeq{\end{equation}}
\def\0{\otimes}
\def\6{\langle}
\def\9{\rangle}
\def\bp{{\bf p}}
\def\br{{\bf r}}

\begin{document}

\begin{center}{\large {\bf
Opposite momenta lead to opposite directions}}\\[8mm]

Asher Peres\\[3mm]
{\it Department of Physics, Technion---Israel Institute of Technology,\\
32000 Haifa, Israel}\\[8mm]

{\bf Abstract}\end{center}

\begin{quote}
When a particle decays into two fragments, the wavefunctions of the
latter are spherical shells with expanding radii. In spite of this
spherical symmetry, the two particles can be detected only in opposite
directions.\end{quote}\vspace{15mm}

\noindent When particles at rest decay into two fragments, the latter
have opposite momenta,

\beq \bp_1+\bp_2=0. \label{sump} \eeq
Yet, the statistical distribution of each fragment is isotropic. The
same holds for pairs of particles resulting from a collision described
in the center-of-momentum frame. In classical physics, it is obvious
that particles with opposite momenta will be found in opposite
directions. The problem is whether this property still holds in quantum
mechanics. Indeed, the operator $\bp_1+\bp_2$ does not commute
with ${\bf q}_1+{\bf q}_2$\,. Rather, there are uncertainty relations,

\beq \Delta(p_1+p_2)\,\Delta(q_1+q_2)\geq\hbar, \label{uncert}\eeq
for each one of the Cartesian components of these vectors. This equation
says that if a system is prepared in such a way that $(p_1+p_2)$ is
sharp, then the mid-point between the two particles has a broad
distribution. However, Eq.~(\ref{uncert}) does not restrict the angular
alignment of the two particles. The purpose of the present article is to
show that the operator equation~(\ref{sump}) leads to an observable
alignment of the detection points of the two particles.

First, consider a simpler problem, namely the motion of a single free
particle. Why does a typical wavepacket move along a straight line? Let
us write the initial wavefunction as a Fourier integral,

\beq \psi(\br,0)=\int f(\bp)\,e^{i\bp\cdot\br/\hbar}\,d\bp.\eeq
After a time $t$, this wavefunction becomes

\beq \psi(\br,t)=\int f(\bp)\,e^{i(\bp\cdot\br-Et)/\hbar}\,d\bp,
 \label{psit}\eeq
where $E=\bp^2/2m$ for nonrelativistic particles (only the
nonrelativistic case is considered here$^1$). Let us write

\beq f(\bp)=|f(\bp)|\,e^{iS(\bp)/\hbar}, \eeq
and let us assume that $|f(\bp)|$ is peaked around $\bp\simeq{\bf k}$,
so that the main contribution to the integral in Eq.~(\ref{psit}) comes
from values of \bp\ in the vicinity of {\bf k}. However, this is not the
only condition on the parameters appearing in that integral. Its value
is usually very small because of the rapid oscillations of the
exponent.  The integral will be appreciably different from zero only if
the phase of the exponent is stationary, namely

\beq {{\partial S}\over{\partial\bp}}+\br
  -{{\partial E}\over{\partial\bp}}\,t=0, \eeq
where all the above expressions have to be evaluated for $\bp={\bf k}$.
Recall that $\partial E/\partial\bp={\bf v}$ is the (classical) velocity
of a particle with momentum \bp, and define $\br_0:=
\partial S/\partial\bp$ (evaluated at $\bp={\bf k}$). We then have

\beq \br=\br_0+{\bf v}\,t.\eeq
We see that for a wavepacket given by Eq.~(\ref{psit}), $\psi(\br,t)$ is
large only in the vicinity of the above value of \br. In other words,
the wavepacket moves in a way similar to that of a classical particle,
provided that $|f(\bp)|$ is indeed peaked near some value of \bp, and
that $S(\bp)$ is well behaved there.

If there are two particles, we write likewise

\beq \psi(\br_1,\br_2,t)=
 \int f(\bp_1,\bp_2)\,e^{i(\bp_1\cdot\br_1+\bp_2\cdot\br_2-Et)/\hbar}\,
  d\bp_1\,d\bp_2. \eeq
Then, if $|f(\bp_1,\bp_2)|$ is peaked around $\bp_1\simeq{\bf k_1}$ and
$\bp_2\simeq{\bf k_2}$, the above $\psi$ describes two wavepackets, with
approximate positions

\beq \br_1=\br_{10}+{\bf v}_1t,\eeq
and

\beq \br_2=\br_{20}+{\bf v}_2t,\eeq
where $\br_{j0}=-\partial S/\partial\bp_j$ and ${\bf v}_j=\partial
E/\partial\bp_j$ (evaluated at $\bp_1={\bf k_1}$ and $\bp_2={\bf k_2}$).

However, this description of two wavepackets moving along straight lines
does not correspond to the physical situation discussed at the beginning
of this article, namely two particles having opposite momenta and an
{\it isotropic distribution\/}. In that case, each wavepacket is an
expanding shell. However, the two spherical shells are correlated, and
it will be shown below that the two particles in each pair are observed
along diametrically opposite directions.

Let $F(\bp_0,E_0)$ be the momentum and energy distribution of the
particle that decays into two fragments (or of the two incoming
particles in a collision). From energy and momentum conservation, we
have

\beq f(\bp_1,\bp_2)=\int F(\bp_0,E_0)\,\delta(\bp_0-\bp_1-\bp_2)\,
  dE_0\,d\bp_0, \eeq
\beq \phantom{f(\bp_1,\bp_2)}=F(\bp_1+\bp_2,E_1+E_2). \eeq
We assume that $F$ is peaked around $\bp_0=0$, so that Eq.~(\ref{sump})
holds for the expectation values of $\bp_1$ and $\bp_2$. Moreover, let
us assume that the phase of $F$ satisfies $\partial S/\partial\bp_j=0$,
so that the decaying system is localized near the origin of the
coordinates. It is always possible to choose the coordinate system so
that these conditions hold. We then have, as before,

\beq \psi(\br_1,\br_2,t)= \int F(\bp_1+\bp_2,E_1+E_2)\,
  e^{i(\bp_1\cdot\br_1+\bp_2\cdot\br_2-Et)/\hbar}\,d\bp_1\,d\bp_2,
 \label{psi12} \eeq
where $E=E_1(p_1)+E_2(p_2)$. As usual, the phase of the exponent has to
be stationary to yield a non-negligible value of $\psi$. However, in the
present case, $F$ is not peaked at definite values of $\bp_1$ and
$\bp_2$, but has spherical symmetry. We therefore introduce spherical
coordinates such as 
\beq p_{1x}=p_1\sin\theta_1\cos\phi_1,\eeq
and

\beq r_{1x}=r_1\sin\theta'_1\cos\phi'_1,\eeq
and likewise for the other components. We thus have

\beq \bp_1\cdot\br_1=p_1\,r_1\,\cos\xi_1, \eeq
where $\xi_1$ is the angle between $\bp_1$ and $\br_1$, explicitly given
by

\beq \cos\xi_1=\cos\theta_1\,\cos\theta'_1+\sin\theta_1\,\sin\theta'_1\,
  \cos(\phi_1-\phi'_1), \eeq
and likewise for the second particle.

For given values of $\br_1$ and $\br_2$, the phase in Eq.~(\ref{psi12})
has to be stationary with respect to variations of the six integration
variables $p_j$, $\theta_j$ and $\phi_j$. Since the various angles
appear in that phase only in the expressions for $\cos\xi_1$ and
$\cos\xi_2$, we can as well request $\cos\xi_j$ to be stationary. This
implies that $\sin\xi_j=0$, or simply $\xi_j=0$. (We also have
$\sin\xi_j=0$ when $\xi_j=\pi$, namely when $\bp_j$ and $\br_j$ have
opposite directions. However, this case corresponds to $t\to-\infty$
and is irrelevant to the present problem.)

The exponent thus becomes $i(p_1r_1+p_2r_2-E_1t-E_2t)/\hbar$ and
stationarity with respect to $p_1$ and $p_2$ leads to

\beq r_1=v_1t\qquad\qquad\mbox{and}\qquad\qquad r_2=v_2t,\eeq
where $v_j=dE_j/dp_j$ is the classical velocity of a particle of energy
$E_j$.

Recall now that $F(\bp_1+\bp_2,E_1+E_2)$ is peaked at $\bp_1+\bp_2=0$
and at $E_1+E_2=E_0$. In spherical coordinates, these peaks occur at

\beq \theta_1+\theta_2=\pi\qquad\qquad\mbox{and}\qquad\qquad
 |\phi_1-\phi_2|=\pi, \label{ang} \eeq
and

\beq p_1{}^2=p_2{}^2=2\mu E_0, \eeq
where $\mu=m_1m_2/(m_1+m_2)$ is the reduced mass of the pair of outgoing
particles.

We have seen above that the demand of a stationary phase in the
integrand in Eq.~(\ref{psi12}) gives $\xi_j=0$. It follows that the
directions of $\br_1$ and $\br_2$ have to satisfy relations similar
to Eq.~(\ref{ang}):

\beq \theta'_1+\theta'_2=\pi\qquad\qquad\mbox{and}\qquad\qquad
|\phi'_1-\phi'_2|=\pi. \eeq
We thus see that opposite momenta lead to opposite directions, as
intuitively expected. A spectacular experimental verification of this
property was recently given by Pittman {\it et al.\/}$^2$

Finally, we have to evaluate how large may be deviations from perfect
alignment. There are two causes for these deviations. One is that each
wavepacket has a width which increases with time (until now we were only
concerned by the motion of its centroid). The expressions
$pr\cos\xi/\hbar$ in the exponent in Eq.~(\ref{psi12}) are maximal for
$\xi=0$ (that is, when \bp\ and \br\ are parallel) and rapid
oscillations start at $pr\xi^2/h\sim1$, or $\xi\sim\sqrt{h/pr}=
\sqrt{\lambda/r}$, where $\lambda$ is the de~Broglie wavelength of the
particles. As expected, the transversal deviation $\sqrt{\lambda r}$ is
identical to the standard quantum limit$^3$ $\sqrt{ht/m}$.

The other cause of deviations from perfect alignment is that $\bp_0=
\bp_1+\bp_2$ is not exactly zero, but has a width $\Delta p_0$. The
resulting angular spread is of the order of $(\Delta p_0)/p_j=
(\Delta p_0)/\sqrt{2\mu E_0}$. This spread does not decrease as
$r\to\infty$, so that it is the main cause of deviations for
$r>hp/(\Delta p_0)^2$.

This work was supported by the Gerard Swope Fund and the Fund for
Encouragement of Research.\\[10mm]

\frenchspacing\begin{description}
\item $^1$For the relativistic case, see A. Peres, Ann. Phys. (NY) {\bf37},
179 (1966).

\item $^2$T. B. Pittman, Y. H. Shih, D. V. Strekalov, and A. V. Sergienko,
Phys. Rev. A {\bf52}, 3429 (1995). This experiment does not involve
massive particles with opposite momenta (as I consider in the present
article), but correlated photons resulting from parametric
down-conversion. However, the reason for their observed spatial
correlation is the same as explained here.

\item $^3$C. M. Caves, Phys. Rev. Letters {\bf54}, 2465 (1985).
\end{description}
\end{document}